\documentclass[aps,prl,twocolumn,groupedaddress, showpacs]{revtex4-1}
\usepackage{graphicx}
\usepackage{color}
\usepackage{amsmath}
\usepackage{latexsym}
\begin{document}

\title{Chaotic Griffiths Phase with Anomalous Lyapunov Spectra in Coupled Map Networks }

\author{Kenji Shinoda}
\email[]{shinoda@complex.c.u-tokyo.ac.jp}
\author{Kunihiko Kaneko}
\email[]{kaneko@complex.c.u-tokyo.ac.jp }
\affiliation{Department of Basic Science, The University of Tokyo, \\
3-8-1 Komaba, Meguro-ku, Tokyo 153-8902, Japan}


\begin{abstract}
Dynamics of coupled chaotic oscillators on a network are studied using coupled maps. Within a broad range of parameter values representing the coupling strength or the degree of elements, the system repeats formation and split of coherent clusters. The distribution of the cluster size follows a power law with the exponent $\alpha$, which changes with the parameter values. The number of positive Lyapunov exponents and their spectra are scaled anomalously with the power of the system size with the exponent $\beta$, which also changes with the parameters. The scaling relation $\alpha \sim 2(\beta +1)$ is uncovered, which seems to be universal independent of parameters and networks.
\end{abstract}


\maketitle

After the success of extensive studies on network structures\cite{watts1998collective, barabasi1999emergence, albert2002statistical}, the dynamics on networks has gathered considerable attention\cite{gross2009adaptive}.  Apart from simple two-state dynamics as adopted in simplified neural-networks or gene-regulatory networks as well as in epidemic propagation, oscillatory dynamics on a network have been extensively investigated. The main focus therein lies in how synchronization is achieved globally among all oscillators and in determining the condition under which this is achieved: Depending on the network structure, synchronizability varies, and the design of such networks structures that are easily synchronized has been mathematically analyzed\cite{nishikawa2006maximum, motter2005network}.

When oscillators are globally synchronized, their dynamics are reduced into just that of a single oscillator. If the dynamic elements on the network are responsible for some function, then, such global synchronization would generally imply the loss of the function. In the power grid network, synchronization will lead to a global black-out\cite{dobson2001initial, buldyrev2010catastrophic}, while in neural networks, global synchronization may lead to the loss of cognitive function.  In contrast, biological systems often avoid such global synchronization, and instead involve dynamics with many degrees, which are often suggested to lie at a critical state, represented by a power-law in activities\cite{nykter2008gene,furusawa2003zipf,furusawa2012adaptation,mora2011biological, shew2009neuronal,chialvo1999learning, markovic2014power}. Hence, there is a need to study the dynamics on the network beyond simple synchronization, which achieves a critical state. 

Indeed, dynamics with many degrees of elements are much richer. They include pattern selection and intermittency, split of elements into multiple coherent clusters, chaotic itinerancy that changes effective degrees of synchronization in time, and collective chaotic dynamics that are distinct from elementary dynamics.  These high-dimensional chaotic dynamics have been extensively studied in coupled map lattices (CMLs)\cite{kaneko1984period,kaneko1989pattern,kaneko1993theory} or globally coupled maps (GCMs)\cite{kaneko1990clustering}, in which simple, identical chaotic dynamics typically represented by a one-dimensional map interact with each other.
Behaviors discovered in coupled maps have been observed in fluid, optical, electronical, and chemical systems as well as in biological and neural dynamics\cite{kaneko2000chaos}, while direct experiments on coupled maps have also been carried out\cite{hagerstrom2012experimental}.

The CML is a dynamical system on a regular lattice, while the GCM adopts all-to-all, mean-field coupling. In considering chaotic dynamics on a network, then, coupled maps on networks (CMNs) should be relevant for exploring the salient behaviors in high-dimensional dynamics. In CMNs, so far, conditions for chaotic synchronization\cite{jost2001spectral,atay2004delays} and splitting of elements into a few synchronized clusters, which partially depends on network structures \cite{manrubia1999mutual,ito2001spontaneous, ito2003spontaneous, jalan2003self, jalan2005synchronized, amritkar2005synchorized, barahona2002synchronization, gade2000synchronous}, have been investigated. Neither complex chaotic dynamics nor a critical state robust to parameter changes, however, has been explored as yet.

In the present Letter, we study a CMN with chaotic logistic maps. After classifying the dynamics into several phases, we focus on a phase that we call chaotic Griffiths phase, in which elements repeat synchronization and desynchronization intermittently.  In this phase, the size distribution of synchronized clusters is found to follow a power law, while the Lyapunov spectra satisfy anomalous scaling. These ``critical" behaviors are maintained over a broad range in the parameter values. Furthermore, the critical exponents of the distribution and Lyapunov spectra change with the parameter values, while maintaining a certain scaling relationship.

To be specific, we study the following coupled map network
\begin{equation}
x_{n+1}(i)=(1-\epsilon)f(x_n(i))+\frac{\epsilon}{k_i} \sum_{j=1}^{N}T_{i,j}f(x_n(j))
\label{eqn:CMN}
\end{equation} with the logistic map $f(x)=1-ax^2$, and the coupling strength $\epsilon$, where the adjacency matrix $T_{i,j}$ represents Erd\"{o}s-R\'{e}nyi random network and $k_i$ is a degree of the element $i$, whose average is set at $k$. Here, we choose a sufficiently large value of $a$ (say $1.6<a<2$), such that chaotic dynamics exist for the logistic map $x_{n+1}=f(x_n)$. As analyzed by linear stability analysis, the whole elements are synchronized for larger $\epsilon$ and $k$. The dynamics of this synchronized state are reduced to a single logistic map $x_{n+1}=f(x_n)$, and thus exhibit chaotic dynamics. With the decrease in $\epsilon$ or $k$, the synchronized state loses the stability, and nontrivial dynamics appear. The attractor dynamics are roughly classified into the following phases, as are also quantitatively characterized by the Lyapunov exponents $\lambda_1 \geq \lambda_2\geq \cdots \geq \lambda_N$ (see Supplemental Fig.1).

(i)Chaotic Synchronization:  All elements are synchronized, and thus, the dynamics are reduced to a single logistic map. $\lambda_1>0$, and $\lambda_j<0$ for $j>1$.
(ii) Chaotic Griffiths Phase (CGP): $\lambda_j>0$ up to certain number $N_p$ with $1\ll N_p \ll N$. As shown in Fig.1a, $x_n(i)$ 's are almost synchronized for some time and are desynchronized later.  Criticality with power-law statistics is preserved within the phase, as will be discussed in detail below.
 (iii) Ordered Phase: After transients, chaos disappears and is replaced by a periodic or quasiperiodic attractor (see Fig.1b), such that $\lambda_1\leq0$. The phase corresponds to the ordered phase in GCM \cite{kaneko1990clustering} or pattern selection in CML\cite{kaneko1989pattern}. Near the boundary to phase (ii), the chaotic transient before reaching the attractor is quite long. 
(iv) Frozen Chaos Phase with Macroscopic Order: Dynamics of each element are desynchronized and chaotic, while maintaining the period-2 band motion as $x>x^* \leftrightarrow x<x^*$, where $x^*$ is the unstable fixed point of the map $x_{n+1}=f(x_n)$ (see Fig.1c). Here, many but not all of the Lyapunov exponents are positive. The number of positive Lyapunov exponents $N_p$ increases linearly with $N$, i.e., $N_p=O(N)<N$. The phase corresponds to the frozen random pattern in CML.
(v) Fully chaotic phase: All the Lyapunov exponents are positive, i.e., $N_p=N$, and dynamics are fully chaotic (see Fig.1d). The phase corresponds to the turbulent state in CML and GCM.

\begin{figure}[]
\includegraphics[width=8cm]{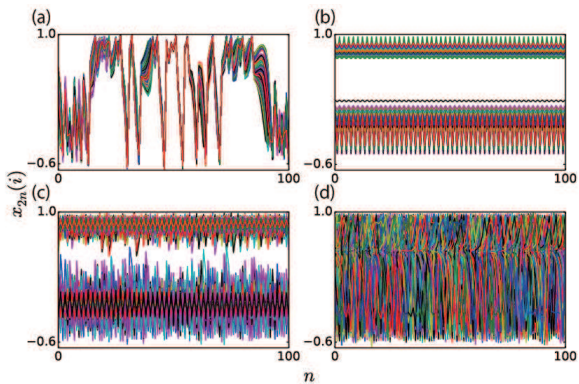}
\caption{Examples of typical time series in the CMN (1): $x_{2n}(i)$ as a function of time step $n$ are overlaid for all elements $i$, after discarding initial transients of $10^6$ steps. Instead of plotting every time step, the variables are plotted every two time steps to make them clearly discernible, since the period-two oscillation is inherent in the logistic map. $a=1.7$ and $N=200$. (a) $\epsilon=0.5$, $k=20$ (phase (ii)). (b) $\epsilon=0.35$, $k=15$ (phase (iii)).  (c) $\epsilon=0.2$, $k=10$ (phase (iv)).  (d) $\epsilon=0.05$, $k=10$ (phase (v)).}
\label{fig1}
\end{figure}

\begin{figure}[]
\includegraphics[width=8cm]{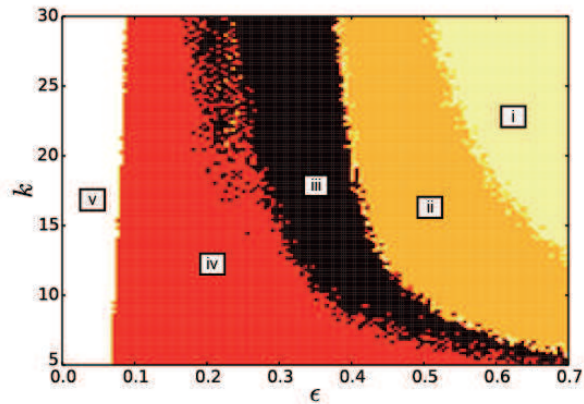}
\caption{Phase diagram of the CMN (1) with $a=1.7$ and $N=200$.  Each phase (i)-(v) (see text) is determined by the Lyapunov exponents as described in the text (see Supplemental Material for the diagrams on $\lambda_1$, $\lambda_2$, $N_p$ and an index for macroscopic order defined therein). The configuration of the phase diagram is independent of $a$, while the phase boundary is shifted.
}
\label{fig2}
\end{figure}

\begin{figure}[]
\includegraphics[width=8cm]{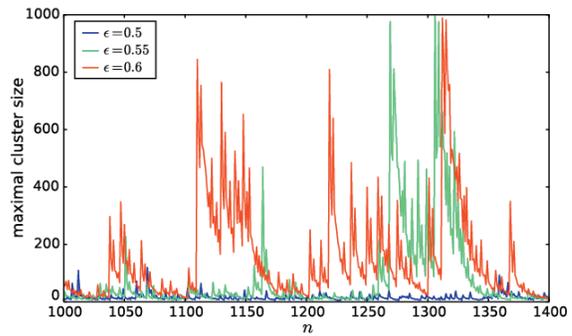}
\caption{Temporal evolution of the maximal cluster size. $a=1.7$, $k=20$, and $N=1000$. The cluster is computed by using the threshold $\delta=10^{-3}$, while this intermittent behavior does not vary as long as it is sufficiently small. $\epsilon=0.5$ (blue line), $\epsilon=0.55$ (green line) $\epsilon=0.6$ (red line), in the chaotic Griffiths phase.
}
\label{fig3}
\end{figure}

Hereafter, we focus on phase (ii) (CGP), as it is inherent to a network system, and there is no correspondent phase in CML and GCM. First, we analyze the repetition of the synchronization--desynchronization process by defining a cluster with a given precision: By introducing bins with the precision $\delta$, which is sufficiently small, we define a cluster as elements $x_n(i)$ that fall within the same bin. In Fig.3, the time series of maximal cluster size is plotted. As shown, many of the elements are synchronized from time to time to form a large cluster, and then, they are desynchronized.  This represents the repetition of the synchronization--desynchronization process. The behavior is analogous with chaotic itinerancy in GCM\cite{kaneko1990clustering,CI}, but a remarkable feature here is that this phase exists not only at a critical point but also in a broad parameter region (see Fig.2). Now, to confirm criticality, we computed the distribution $P(s)$ of cluster sizes $s$ by long-term sampling of them. As shown in Fig.4, the distribution obeys a power law $P(s)\sim s^{-\alpha}$ at this phase, where the exponent $\alpha$ changes with the parameters $\epsilon$,$k$, and $a$. For given $a$, as $\epsilon$ or $k$ is increased to approach the chaotic synchronization phase (i), the exponent $\alpha$ approaches 2, while it increases monotonically to $\sim 4$ as the parameters decrease towards the boundary value to phase (iii).  This suggests that the criticality is maintained throughout phase (ii), while the exponent $\alpha$ changes monotonically with the parameters. (As for the change against $k$, see Supplemental Fig.2).

\begin{figure}[]
\includegraphics[width=8cm]{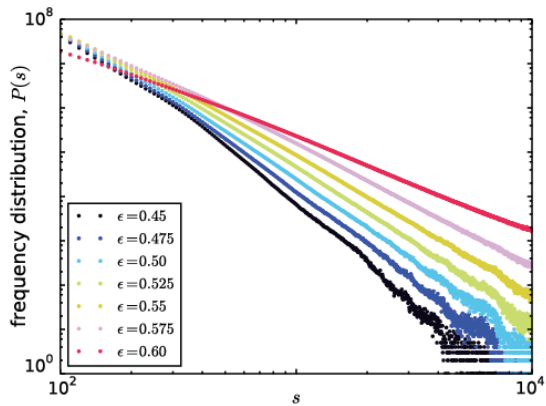}
\caption{The distribution $P(s)$ of cluster size $s$. Log--log plot. $a=1.7$, $k=20$, and $N=16384$. The results from $\epsilon=$ 0.45, 0.475, 0.5, 0.525, 0.55, 0.575, and 0.6 are plotted with different colors. The distribution is obtained by sampling over $10^{3}$ steps, with $100$ initial conditions, over 100 networks, by using the threshold $\delta=10^{-3}$, while the exponents do not vary as long as this threshold is sufficiently small, and also the network sample dependence is negligible. }
\label{fig4}
\end{figure}

\begin{figure}[]
\includegraphics[width=8cm]{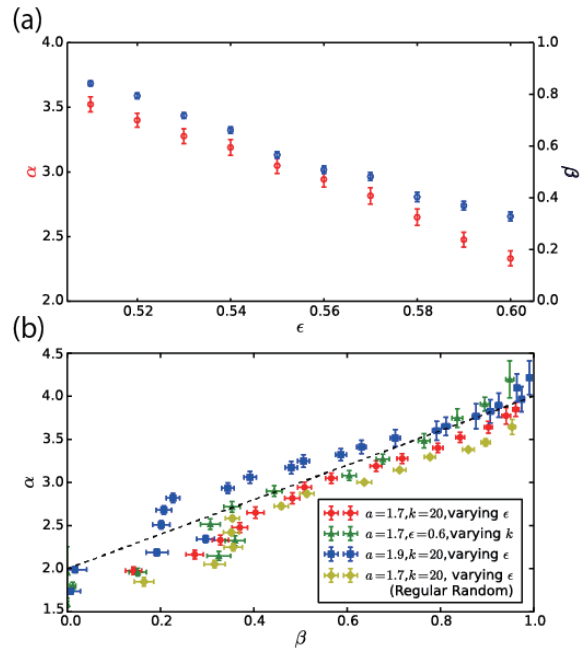}
\caption{(a) Dependence of the exponent $\alpha$ and $\beta$ upon $\epsilon$, from the data of Fig.4 and Fig.6.
(b) Relationship between the exponents $\alpha$ (abscissa) and $\beta$(ordinate). Apart from the data of Fig.4 ($\circ$), the data with varying $k$ by fixing $\epsilon$ at 0.6 ($\triangle$), the data $a=1.9$ ($\Box$), and a=1.7 on a regular random network ($\Diamond$) with fixed $k$ and varying $\epsilon$ are plotted. The error bars are computed from the least square fit of the power law.}
\label{fig5}
\end{figure}

\begin{figure}[]
\includegraphics[width=8cm]{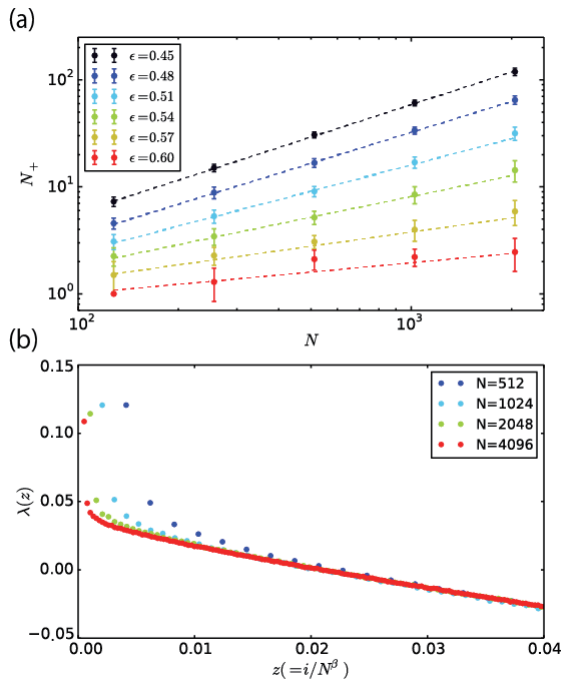}
\caption{(a) The number of positive Lyapunov exponents $N_p$ plotted as a function of the system size $N$, for $\epsilon=$ 0.45, 0.48, 0,51, 0.54, 0.57, and 0.6 with different colors. $a=1.7$, $k=20$.
 (b) The scaled Lyapunov spectra $\lambda(z)$ with $z=i/N^{\beta}$. The Lyapunov exponents are computed from $2000$ time steps over 12 networks samples and 4 initial conditions.
}
\label{fig6}
\end{figure}

Existence of power-law behavior over a finite range of parameters is not typically observed in a regular lattice or a mean-field coupling system. Indeed, in the spatiotemporal chaos in CML, the correlation decays exponentially in space\cite{bohr1989size}, while in the turbulent phase in GCM, the distribution of synchronized cluster decays exponentially such that a large cluster is not generated\cite{kaneko1990clustering}. In these cases, the number of positive Lyapunov exponents $N_p$ increases linearly with the number of elements $N$, i.e., they are extensive variables\cite{kaneko1986lyapunov,kaneko1994information,nakagawa1995anomalous,takeuchi2011extensive}. In contrast, in phase (ii) with a power-law behavior, the number of positive Lyapunov exponents $N_p$ increases with $N$ anomalously, as $N^{\beta}$ (see Fig.6a).
The exponent is $\sim 1$ at the boundary with phase (iii), and decreases monotonically as $\epsilon$ or $k$ is increased, until it approaches 0 at the boundary with phase (i). 
Furthermore, the Lyapunov exponents $\lambda(z)$ plotted as a function of the scaled variable $z=i/N^{\beta}$ follow a single curve independent of $N$ for positive exponents, except for the first few Lyapunov exponents (see Fig.6b and Supplemental Fig.3).
This anomalous scaling is in stark contrast with the Lyapunov exponents in CML and GCM, where only the normal scaling $\beta=1$ is observed (except at the critical point)\cite{Note-Lyap}.

The exponent $\beta<1$ in the CMN implies that the degree of chaos does not properly increase with the system size, thus allowing room for generation of a large coherent cluster intermittently. Then, it is expected that with the increase from $\beta =0$ to 1, the fraction of a larger cluster decreases, such that the exponent $\alpha$ is increased.  The relationship between the two exponents is plotted in Fig.5, where the data are roughly fitted with $\alpha \sim 2(\beta+1)$.

This critical behavior within a finite parameter range in a system with quenched randomness is reminiscent of Griffiths phase, first predicted in the diluted Ising model\cite{griffiths1969nonanalytic}. The existence of Griffiths phase in network dynamics was recently reported\cite{munoz2010griffiths, vazquez2011temporal, moretti2013griffiths}: Mu{\~n}oz et al. studied a quenched contact process to find a power-law relaxation process within a finite range of remaining rate of edges. Indeed, in a contact process on a regular lattice, there exists a percolation threshold beyond which active states persist, and below which active states disappear to be replaced by the absorbing state that is reached within a finite time.  For a quenched disordered system, in contrast, the transition point is blurred, and the critical behavior is stretched in the parameter space, leading to power-law relaxation in time to the absorbing state.

In GCM, similarly, there is a transition to chaotic synchronization at a certain coupling strength $\epsilon$, and in the present CMN, this transition is blurred, leading to perpetual repeat of formation and collapse of synchronized clusters, resulting in its power-law distribution. In this sense, phase (ii) corresponds to the Griffiths phase. We should note, however, that the global synchronization is not an absorbing state, and the power-law behavior exists as a chaotic attractor, not in the relaxation process, in contrast to the network Griffiths phases studied so far. Hence, we call regime (ii) as ``chaotic Griffiths phase".

The scaling relationship $\alpha\sim 2(\beta+1)$ seems to be valid, independent of $\epsilon$ and $k$, as well as of $a$\cite{a-dep}.
Theory for it is not yet developed, while a rough argument is given as follows.  In the limit to the boundary to chaotic synchronization, $\beta \rightarrow 0$ and $\alpha\rightarrow 2$. Now for simple approximation, let us represent the change in cluster size $s$ by a random walk of $s$. In this case, as the size $s$ increases the probability of move at each step increases linearly with $s$, as each element in the cluster can synchronize or desynchronize with others.
Then the stationary distribution approaches $P(s)\sim s^{-2}$, as is derived from the Fokker-Planck equation $\partial P(s,t)/\partial t=\partial^2 s^2 P(s,t)/\partial s^2$ corresponding to the stochastic differentiation $ds=s\circ dt$ of Ito calculus.
Next, when $\beta>0$, the probability of move is expected to increase with the cluster size $s$ as $s^{1+\beta}$, since the degrees for chaos in $s$ elements increase with the power $\beta$, leading to additional increase of move probability with $s^{\beta}$.
Thus, the above equation would be changed to $ds=s^{1+\beta}\circ dt$. The stationary distribution of the corresponding Fokker-Planck equation is then given by $\sim s^{-2(1+\beta)}$, leading to $\alpha \sim 2(1+\beta)$. This argument, of course, is rather rough, and needs to be elaborated by complete theoretical explanation in the future\cite{coagulation}.

Apart from the present Erd\"{o}s-R\'{e}nyi random network of coupled maps, we have also investigated CMN of some other topologies, such as regular random and  small-world networks.  Chaotic Griffiths phase is generally observed when the global synchronization is lost. Interestingly, the relationship between exponents $\alpha$ and $\beta$ seems to be still valid (see Fig.5b for the relationship for regular random networks) \cite{scale-free}. 

To sum up, we have reported chaotic Griffiths phase in CMN: elements repeat formation of synchronization and desynchronization, where the size of synchronized clusters follows the power-law distribution with the power $\alpha$.  The exponent $\alpha$ changes monotonically in the phase, where the Lyapunov exponents follow the anomalous scaling with another exponent $\beta$, while roughly maintaining a monotonic relationship with $\alpha \sim 2(\beta+1)$. This relationship is independent of the network structure: Dynamics with desynchronization of coherence by chaos dominates the network structure.  Theory for the scaling relationship between $\alpha$ and $\beta$ has to be developed in the future, possibly with the aid of renormalization group.

The prevalence of critical states is extensively reported in biological networks, especially in brain dynamics as a correlation of neural activities\cite{chialvo1999learning,shew2009neuronal, petermann2009spontaneous,chialvo2010emergent}. The module structure in neural connectivity is often explored as the origin of criticality. The present chaotic Griffiths phase may provide an alternative view on this, considering that chaotic neural dynamics are often reported.

\begin{acknowledgments}
We would like to thank Dr. Nen Saito and Koji Hukushima for the stimulating discussions.
\end{acknowledgments}

\pagebreak
{\Large{\bf SUPPLEMENTAL MATERIALS}}

\begin{figure}[h]
  \begin{minipage}{1.0\hsize}
 \begin{center}
\includegraphics[width=10cm]{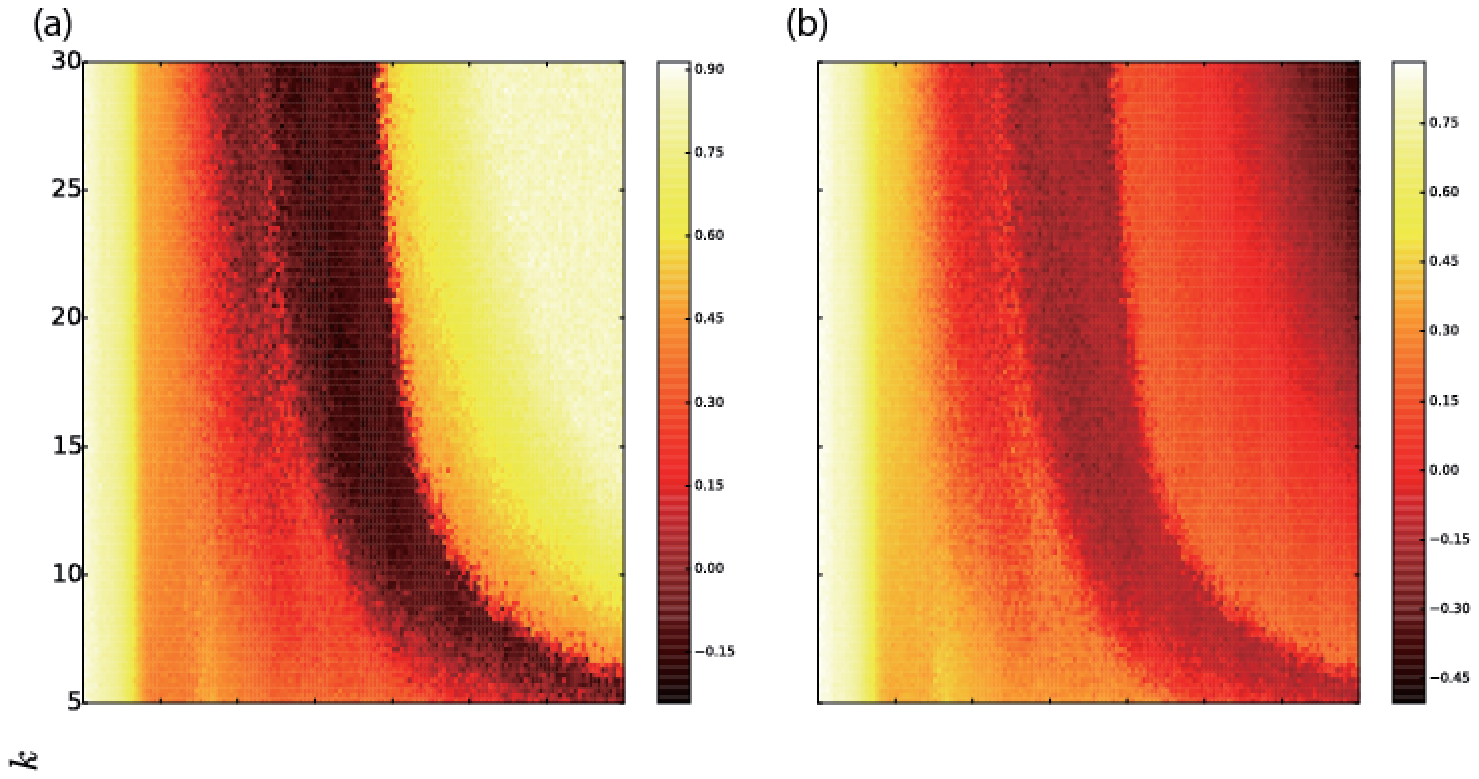}
 \end{center}
 \end{minipage}
 \\
 \begin{minipage}{1.0\hsize}
  \begin{center}
\includegraphics[width=10cm]{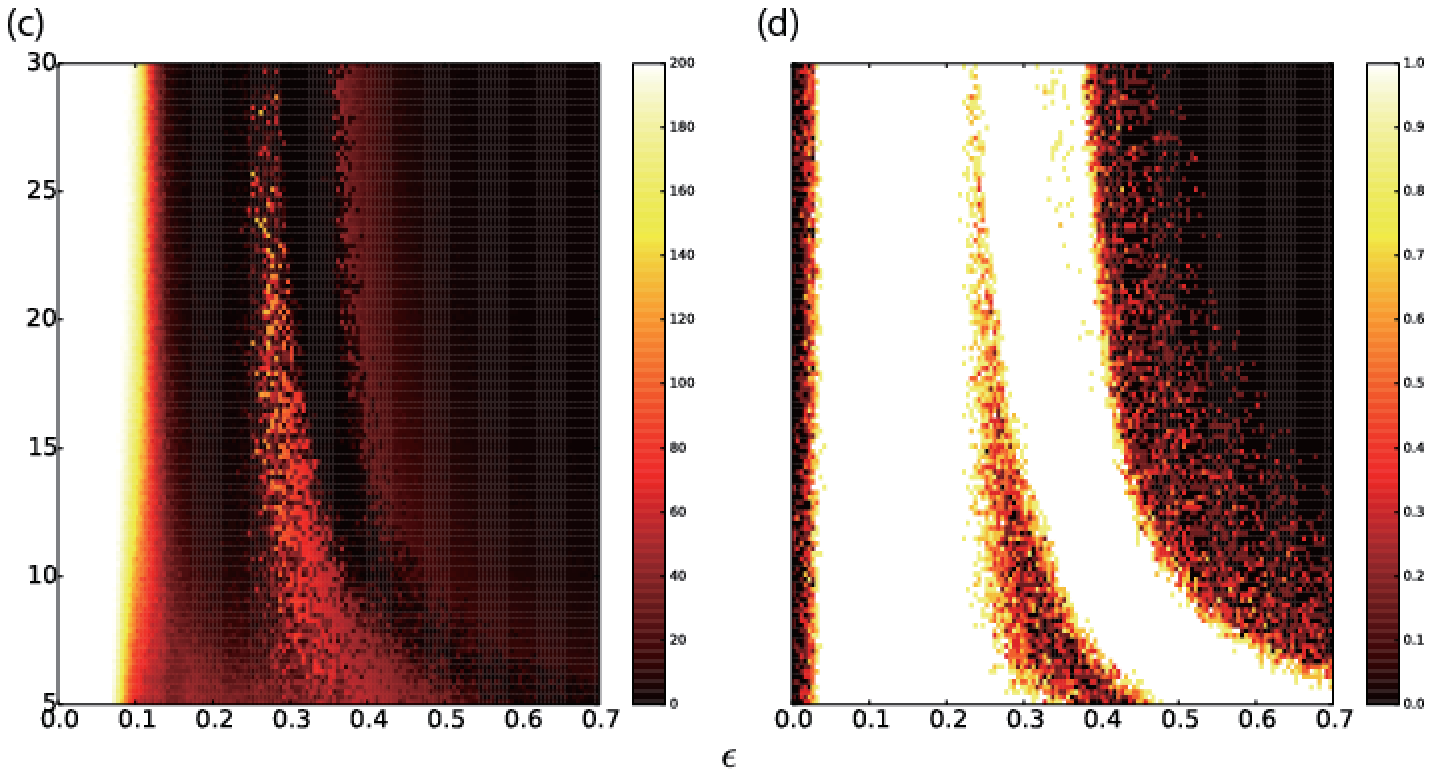}
  \end{center}
 \end{minipage}
\begin{quote}
{\bf SF1} Heatmaps of the values of Lyapunov exponents and the index of the macroscopic order, for the CMN with $a=1.7$ and $N=200$, plotted against $\epsilon$ and $k$. (a) The first Lyapunov exponent, $\lambda_1$. (b) The second Lyapunov exponent, $\lambda_2$. (c) The number of the positive Lyapunov exponents, $N_p$.  (d) The index of the macroscopic order, $Q$. Here, $Q$ is defined as follows: After the dynamics reach an attractor, elements are classified into two groups according to $x_{2n}(i)>x^*$ or $x_{2n}(i)<x^*$ are, where $x^*$ is the unstable fixed point of the logistic map after the relaxation of the system. The  elements $i$ that preserve this inequality are regarded to belong a two-band cluster. $Q$ is the fraction of such elements falling in a two-band cluster. Here, the red or black area around $\epsilon =0.3$ or $0.4$ finally disappears, where elements fall into a tw-cluster after long transients. \end{quote}
\end{figure}

\clearpage
{\Large{\bf SUPPLEMENTAL MATERIALS}}
\begin{figure}[h]
\includegraphics[width=14cm]{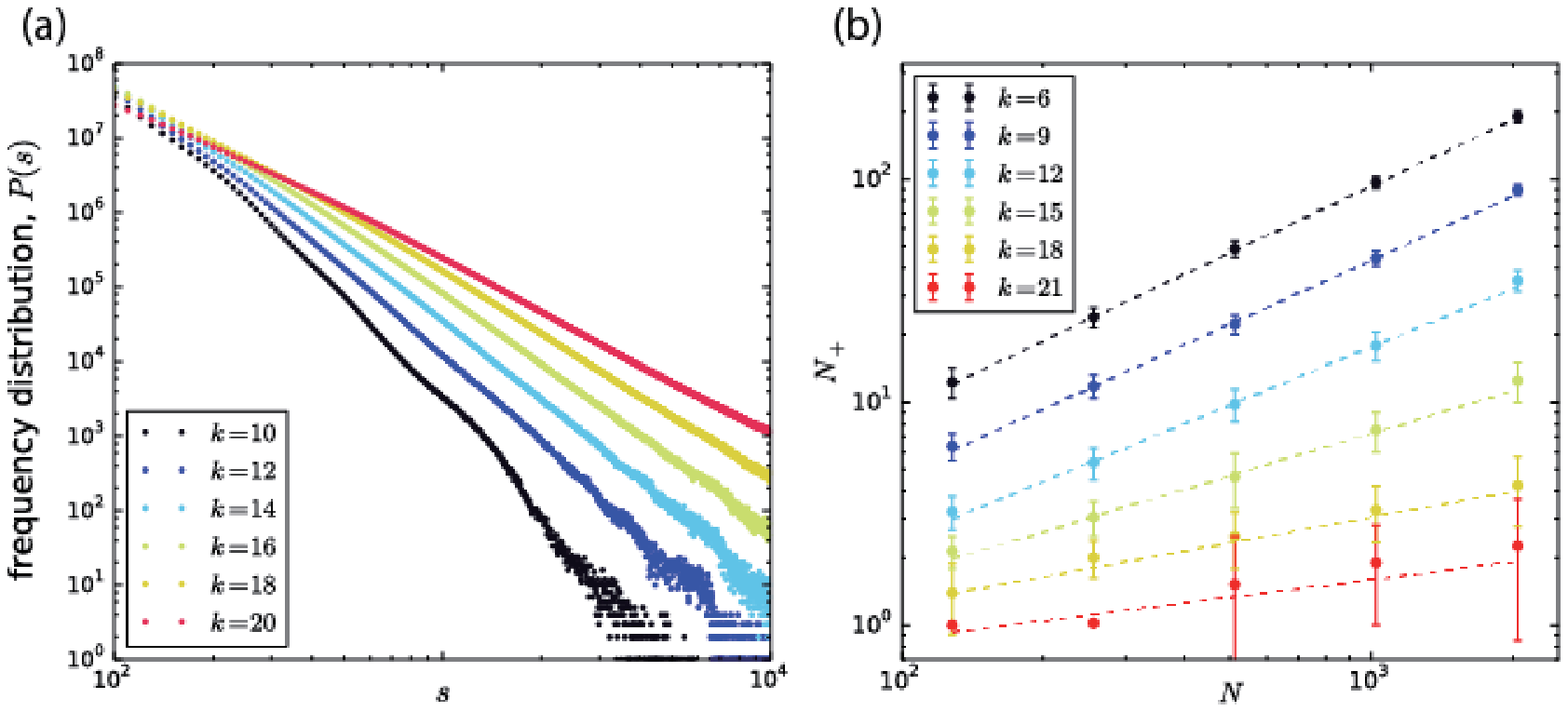}
\begin{quote}
{\bf SF2} (a) The distribution $P(s)$ of cluster size $s$ with varying $k$ by fixing $\epsilon$ at $0.6$. $a=1.7$ and $N=16384$. The results from $k=10, 12, 14, 16, 18, 20$ are plotted with different colors. The distribution is obtained by sampling over $10^{3}$ steps, with $100$ initial conditions, over $100$ networks, by using the threshold $\delta=10^{-3}$, while the exponents do not vary as long as this threshold is sufficiently small. (b) The number of positive Lyapunov exponents $N_p$ plotted as a function of the system size $N$, for $k=6, 9, 12, 15, 18, 21$ with different colors. $a=1.7$, $\epsilon=0.6$.
\end{quote}
\end{figure}

\clearpage
{\Large{\bf SUPPLEMENTAL MATERIALS}}
\begin{figure}[h]
\hspace{0cm}
\includegraphics[width=9cm]{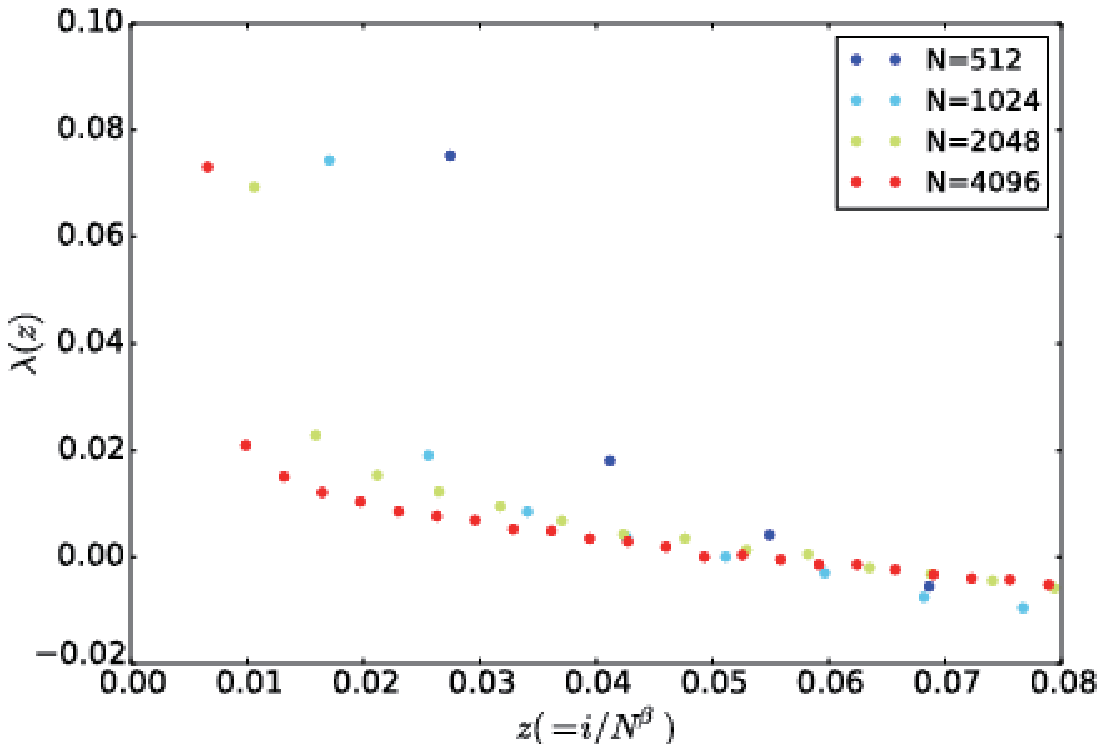}
\vspace{0cm}
\begin{quote}
{\bf SF3} The scaled Lyapunov exponent, plotted as a function of the first Lyapunov exponent with $z=i/N^{\beta}$, $\beta=0.687$ by $k=20 $ and $\epsilon=0.55$. The first exponent $\lambda_1$ is not plotted, which is much larger and close to the value of the Lyapunov exponent of the single logistic map. 
\end{quote}
\end{figure}

\begin{thebibliography}{99}

\bibitem{watts1998collective}
Watts, D.J. and Strogatz, S.H, 1998. Collective dynamics of Small-world networks, Nature, 393(6684), p.440-442. 

\bibitem{barabasi1999emergence}
Barab{\'a}si, A.L. and Albert, R., 1999. Emergence of scaling in random networks, Science, 286(5439), p. 509-512. 

\bibitem{albert2002statistical}
Albert, R. and Barab{\'a}si, A.L., 2002. Statistical mechanics of complex networks, Reviews of modern physics, 74(1), p.47. 

\bibitem{gross2009adaptive}
Gross, T. and Sayama, H., 2009. {\sl Adaptive networks}. Springer Berlin Heidelberg.

\bibitem{nishikawa2006maximum}
Nishikawa, T. and Motter, A.E., 2006. Maximum performance at minimum cost in network synchronization. Physica D: Nonlinear Phenomena, 224(1), pp.77-89.

\bibitem{motter2005network}
Motter, A.E., Zhou, C. and Kurths, J., 2005. Network synchronization, diffusion, and the paradox of heterogeneity. Physical Review E, 71(1), p.016116.



\bibitem{dobson2001initial}
Dobson, I., Carreras, B., Lynch, V. and Newman, D., 2001, An initial model for complex dynamics in electric power system blackouts. In hicss (p. 2017). IEEE.
\bibitem{buldyrev2010catastrophic}
Buldyrev, S.V., Parshani, R., Paul, G., Stanley, H.E. and Havlin, S., 2010. Catastrophic cascade of failures in interdependent networks. Nature, 464(7291), pp.1025-1028
\bibitem{nykter2008gene}
Nykter, M. and Price, N.D. and Aldana, M. and Ramsey, S.A. and Kauffman, S.A. and Hood, L.E. and Yli-Harja, O. and Shmulevich, I., 2008. Gene expression dynamics in the macrophage exhibit criticality, Proceedings of the National Academy of Sciences, 105(6), p.1897-1900.

\bibitem{furusawa2003zipf}
Furusawa, C. and Kaneko, K., 2003. Zipf's law in gene expression, Physical review letters, 90(8), p.088102. 

\bibitem{furusawa2012adaptation}
Furusawa, C. and Kaneko, K., 2012. Adaptation to optimal cell growth through self-organized criticality, Physical review letters, 108(2), p.208103.

\bibitem{mora2011biological}
Mora, T. and Bialek, W., 2011. Are biological systems poised at criticality?, Journal of Statistical Physics, 144(2), p.268-302. 

\bibitem{shew2009neuronal}
Shew, W.L. and Yang, H. and Petermann, T. and Roy, R. and Plenz, D., 2009. Neuronal avalanches imply maximum dynamic range in cortical networks at criticality, The Journal of Neuroscience, 29(30), p15595-15600. 

\bibitem{chialvo1999learning}
Chialvo, D.R. and Bak, P., 1999. Learning from mistakes, Physical Review Letters. 90(4), p.1137-1148. 

\bibitem{markovic2014power}
Markovi{\'c}, D. and Gros, C., 2014. Power laws and self-organized criticality in theory and nature. Physics Reports, 536(2), pp.41-74.

\bibitem{kaneko1984period}
Kaneko, K.,1984. Period-Doubling of Kink-Antikink Patterns, Quasiperiodicity in Antiferro-Like Structures and Spatial Intermittency in Coupled Logistic Lattice - Towards a Prelude of a Field Theory of Chaos,  Progress of Theoretical Physics, 72(3), p.480-486. 

\bibitem{kaneko1989pattern}
Kaneko, K.,1989. Pattern dynamics in spatiotemporal chaos: Pattern selection, diffusion of defect and pattern competition intermittency, Physica D: Nonlinear Phenomena, 34(1-2), p.1-41. 

\bibitem{kaneko1993theory}
Kaneko, K., 1993. {\sl Theory and applications of coupled map lattices}. John Wiley and Son Ltd.

\bibitem{kaneko2000chaos}
Kaneko, K. and Tsuda, I., 2000. {\sl Chaos and beyond}, Springer, Berlin, 10, p.24. 

\bibitem{hagerstrom2012experimental}
Hagerstrom, A.M. and Murphy, T.E. and Roy, R. and H{\"o}vel, P. and Omelchenko, I. and Sch{\"o}ll, E., 2012. Experimental observation of chimeras in coupled-map lattices, Nature Physics, 8(9), p.658-661. 

\bibitem{kaneko1990clustering}
Kaneko, K., 1990. Clustering, coding, switching, hierarchical ordering, and control in a network of chaotic elements, Physica D: Nonlinear Phenomena, 41(2), p137-172. 

\bibitem{jost2001spectral}
Jost, J. and Joy, M.P., 2001. Spectral properties and synchronization in coupled map lattices. Physical Review E, 65(1), p.016201.

\bibitem{atay2004delays}
Atay, F.M., Jost, J. and Wende, A., 2004. Delays, connection topology, and synchronization of coupled chaotic maps. Physical Review Letters, 92(14), p.144101.


\bibitem{manrubia1999mutual}
Manrubia, S.C. and Mikhailov, A.S., 1999. Mutual synchronization and clustering in randomly coupled chaotic dynamical networks, Physical Review E, 60(2), p.1579. 

\bibitem{ito2001spontaneous}
Ito, J. and Kaneko, K., 2001. Spontaneous structure formation in a network of chaotic units with variable connection strengths. Physical Review Letters, 88(2), p.028701.

\bibitem{ito2003spontaneous}
Ito, J. and Kaneko, K., 2003. Spontaneous structure formation in a network of dynamic elements. Physical Review E, 67(4), p.046226.

\bibitem{jalan2003self}
Jalan, S. and Amritkar, R.E., 2003. Self-organized and driven phase synchronization in coupled maps. Physical Review Letters, 90(1), p.014101.

\bibitem{jalan2005synchronized}
Jalan, S., Amritkar, R.E. and Hu, C.K., 2005. Synchronized clusters in coupled map networks. I. Numerical studies. Physical Review E, 72(1), p.016211.

\bibitem{amritkar2005synchorized}
Amritkar, R.E. and Jalan, S. and Hu, C.K., 2005. Synchronized clusters in coupled map networks. II. Stability analysis, Physical Review E, 72(1), p.016211. 

\bibitem{barahona2002synchronization}
Barahona, M. and Pecora, L.M., 2002. Synchronization in small-world systems, Physical Review Letters, 89(5), p.054101. 

\bibitem{gade2000synchronous}
Gade, P.M. and Hu, C.K., 2000. Synchronous chaos in coupled map lattices with small-world interactions. Physical Review E, 62(5), p.6409.

\bibitem{bohr1989size}
Bohr, T. and Christensen, O.B., 1989. Size dependence, coherence, and scaling in turbulent coupled-map lattices, Physical Review Letters, 63(20), p.2161.

\bibitem{Note-Lyap} For $\lambda_i<0$, as the index $i$ goes larger and is the order of $N$, the anomalous scaling is replaced by the normal scaling $i/N$. As for the KS entropy, it is expected that the anomalous scaling with the exponent $\beta$ is valid. Still, much larger $N$ is needed to reduce the deviation due to the initial few exponents, which are much larger.

\bibitem{griffiths1969nonanalytic}
Griffiths, R.B., 1969. Nonanalytic behavior above the critical point in a random Ising ferromagnet, Physical Review Letters, 23(1), p.17. 

\bibitem{munoz2010griffiths}
Mu{\~n}oz, M.A. and Juh{\'a}sz, R. and Castellano, C. and {\'O}dor, G., 2010. Griffiths phases on complex networks, Physical Review Letters, 105(12), p.128701. 

\bibitem{vazquez2011temporal}
Vazquez, F. and Bonachela, J.A. and L{\'o}pez, C. and Mu{\~n}oz, M.A, 2011. Temporal Griffiths phases, Physical Review Letters, 106(23), p.235702. 

\bibitem{moretti2013griffiths}
Moretti, P. and Mu{\~n}oz, M.A., 2003. Griffiths phases and the stretching of criticality in brain networks, Nature Communications, 4. 
\bibitem{CI}
Kaneko, K. and Tsuda, I., 2003. Chaotic itinerancy. Chaos: An Interdisciplinary Journal of Nonlinear Science, 13(3), pp.926-936.
 
\bibitem{kaneko1986lyapunov}
Kaneko, K., 1986. Lyapunov analysis and information flow in coupled map lattices, Physica D: Nonlinear Phenomena, 23(1-3), p.436-447. 

\bibitem{kaneko1994information}
Kaneko, K., 1994. Information cascade with marginal stability in a network of chaotic elements, Physica D: Nonlinear Phenomena, 77(4), p.456-472. 

\bibitem{nakagawa1995anomalous}
Nakagawa, N. and Kuramoto, Y., 1995. Anomalous Lyapunov spectrum in globally coupled oscillators, Physica D: Nonlinear Phenomena, 80(3), p.307-316. 

\bibitem{takeuchi2011extensive}
Takeuchi, K.A. and Chat{\'e}, H. and Ginelli, F. and Politi, A. and Torcini, A., 2011. Extensive and subextensive chaos in globally coupled dynamical systems, Physical Review Letters, 107(12), p.124101. 

\bibitem{a-dep}
In Fig.5, the data for $a=1.9$ might look slightly deviated from the line $\alpha =2(\beta+1)$. This deviation, however, would be due to the finite size effect in $N$. Indeed the data from smaller $N$ showed lareger deviation, and as $N$ is increased, they slowly  approach the line.

\bibitem{coagulation}
Another possible formulation would be the use of coagulation  and fragmentation of synchronized clusters\cite{vicsek1984dynamic, family1986kinetics, sorensen1987cluster}.

\bibitem{vicsek1984dynamic}
Vicsek, T. and Family, F., 1984. Dynamic scaling for aggregation of clusters. Physical review letters, 52(19), p.1669.

\bibitem{family1986kinetics}
Family, F., Meakin, P. and Deutch, J.M., 1986. Kinetics of coagulation with fragmentation: scaling behavior and fluctuations. Physical review letters, 57(6), p.727.

\bibitem{sorensen1987cluster}
Sorensen, C.M., Zhang, H.X. and Taylor, T.W., 1987. Cluster-size evolution in a coagulation-fragmentation system. Physical review letters, 59(3), p.363.

\bibitem{scale-free}
In the case of scale-free network\cite{barabasi1999emergence}, the relationship between the two exponents is deviated. This is possibly because, the power-law distribution in the connectivity gives additional contribution to the power of the cluster distribution, so that the exponent $\alpha$ is increased.

\bibitem{petermann2009spontaneous}
Petermann, T. and Thiagarajan, T.C. and Lebedev, M.A. and Nicolelis, M.A.L and Chialvo, D.R. and Plenz, D., 2009. Spontaneous cortical activity in awake monkeys composed of neuronal avalanches, Proceedings of the National Academy of Sciences, 106(37), p.15921-15926. 

\bibitem{chialvo2010emergent}
Chialvo, Dante R, 2010. Emergent complex neural dynamics, Nature Physics, 6(10), p.744-750. 

\end{thebibliography}
\end{document}